\documentclass[
    twocolumn,
floatfix,
amssymb,
amsmath,
aps,
prl,showpacs]{revtex4}

\usepackage{times}
\usepackage{graphicx}

\newcommand{\uwa}{School of Physics, The University of Western Australia\protect\\ Nedlands, Western Australia 6907, Australia}

\newcommand{\reference}[1]{
	\bibitem
	{#1}
}

\newcommand{\Iav}{\ensuremath{I_\mathrm{av}}}
\newcommand{\Ep}{\ensuremath{E_\mathrm{pulse}}}

\newcommand{\dd}{\;\mathrm{d}}
\newcommand{\Lint}[1]{\int_{-L/2}^{L/2}#1\dd z}
\newcommand{\micron}{\ensuremath{\mu\mathrm{m}}}
\newcommand{\lam}{\lambda}

\newcommand{\apx}{\ensuremath{\approx}}
\newcommand{\Pd}{\ensuremath{P_{\mathrm{det}}}}
\newcommand{\Pav}{\ensuremath{P_{\mathrm{av}}}}
\newcommand{\tec}{technique}
\newcommand{\vb}{vacuum birefringence}
\newcommand{\ti}{\ensuremath{\tau_\mathrm{int}}}
\newcommand{\npar}{\ensuremath{n_{\|}}}
\newcommand{\nperp}{\ensuremath{n_{\perp}}}
\newcommand{\fpar}{\ensuremath{\nu_{\|}}}
\newcommand{\fperp}{\ensuremath{\nu_{\perp}}}
\newcommand{\lpar}{\ensuremath{l_{\|}}}
\newcommand{\lperp}{\ensuremath{l_{\perp}}}
\newcommand{\unit}[1]{\ensuremath{\,\mathrm{#1}}}
\newcommand{\ee}[1]{\ensuremath{\times 10^{ #1}}}
\newcommand{\brac}[1]{\left( #1 \right)}

\newcommand{\eq}[2]{\begin{eqnarray}#2\label{#1}\end{eqnarray}}
\newcommand{\makefig}[3]{
	\begin{figure}[bt]
	\centering
	\includegraphics[width=3.4in]{#1}
	\caption{#3}
	\label{#2}
	\end{figure}
}

\begin{document}
\bibliographystyle{apsrev}
\title{Detection of Vacuum Birefringence with Intense Laser Pulses}
\author{Andre N. Luiten}
\email{andre@physics.uwa.edu.au}
\author{Jesse C. Petersen}
\altaffiliation{Now with Department of Physics, Simon Fraser University, Burnaby, British Columbia, Canada.}
\affiliation{\uwa}
\date{\today}
\begin{abstract}
We propose a novel technique that promises hope of being the first to directly  detect   a polarization in the quantum electrodynamic (QED) vacuum. The technique is based upon the use of ultra-short pulses of light circulating in low dispersion optical resonators.   We show that the technique circumvents the need for large-scale liquid helium cooled magnets, and more importantly avoids the experimental pitfalls that plague  existing experiments that make use of these magnets. Likely improvements in the performance of optics and lasers would result in the ability to observe vacuum polarization in an experiment of only a few hours duration.

\end{abstract}
\maketitle


It was predicted almost seventy years ago that virtual positron-electron pairs in the quantum electrodynamic vacuum provide a means for interaction  between photons~\cite{euler, heisenberg, weisskopf,schwinger}. As yet this effect has not been observed directly in any laboratory experiment although high energy scattering experiments have shown indirect evidence of  such interactions~\cite{wilson,burke}.  It is believed that these processes play an important role in  extreme astrophysical environments~\cite{heyl}. 

The QED mediated interaction predicts  scattering of real photons from virtual photons in an electromagnetic field as well as direct photon-photon scattering mechanism.  For achievable fields, the principal effect of these processes is  a polarization dependent change of the phase velocity of the interacting photons,  in other words, a laser beam propagating through a region of strong electromagnetic field will observe a birefringent and refractive vacuum, in which both polarization states have a phase velocity that differs from that in a field-free vacuum.  
%
For example, light traversing  a  transverse magnetic field region will experience differing refractive indices for polarization parallel and perpendicular to the magnetic field, with the  birefringence  being of the order of ${\Delta n \sim 10^{-21}}$ for realistic  laboratory magnetic fields ($5 - 10$\,T).  

The conventional route to searching for such field-induced birefringence is to couple intense magnetic fields, produced by  superconducting coils, to an optical delay line or a Fabry-P\'erot resonator which forces the light to traverse the strong field many times~\cite{cameron,pvlas1,bmv,lee}.  In contrast, here we propose a novel method for the experimental detection of \vb\, which  dispenses with the need for any large static magnetic fields, and instead makes use of a combination of frequency-stabilized mode-locked lasers~\cite{diddams,jones-lock} and low dispersion optical resonators~\cite{jones-dco,me-pulse}. 
We propose a search for the vacuum birefringence induced by the extremely high fields that exist within a focussed femtosecond duration pulse of light.  This technique holds the promise of  improved sensitivity while   using only room-temperature apparatus that is both reliable and relatively inexpensive. 

$\Re_{the man} x^{2} \frac{1}{2}$ 
To produce  \vb\ with an optical field, a linearly polarized `pump' beam must interact with a counter-propagating `detection' beam.  
 The refractive indices of the vacuum for detection light polarized parallel and perpendicular to the polarization of the pump beam are denoted as \npar\ and \nperp, and given by \cite{alek,vpol}:
\eq{nn}{\npar=1+\frac{16}{45}\frac{\alpha^2U}{U_e};\;\;\;\;\;\nperp=1+\frac{28}{45}\frac{\alpha^2U}{U_e},}
where $\alpha$ is the fine structure constant, $U$ is the energy density in the optical field and
${U_e={m_e^4c^5}/{\hbar^3}\approx 1.44\times 10^{24}}$
is the Compton energy density of the electron ($m_e$ is the electron rest mass).  From Eq.~(\ref{nn}) we observe a birefringent vacuum of magnitude:
\eq{b}{\Delta n=\frac{4}{15}\frac{\alpha^2U}{U_e}=\frac{4}{15}\frac{\alpha^2\Iav}{c U_e}.}
where \Iav\ is the intensity of the pump field.  These expressions are valid for  infinite plane waves, although they give a birefringence of the correct order for laser beams when the beams are well-collimated over the interaction region.  

In this letter we propose the use of short pulses of intense laser  radiation to generate the high fields necessary to polarize the vacuum. It is by this technique that one can generate average field intensities comparable with those produced by   superconducting magnets. As an extreme example, the peak magnetic field within a 1\,J, 50\,fs pulse that is focussed into  1 $\mu$m$^2$    can be of the order of $10^{5}$\,T~\cite{lee}.  The high confinement of the optical field means that while the peak fields are very high, the total energy stored in the field is much smaller than a static magnetic field that produces an equivalent birefringence signal. The pulsed light technique thus has twin benefits:   it  eliminates large forces from the experiment, and also  make shielding of the detection apparatus from stray fields very simple. In existing searches for vacuum birefringence, spurious signals arising from stray magnetic field modifying the detection system components, or stray forces moving  the detection system components, are responsible for limiting the sensitivity~\cite{lee,bmv,cameron,itnoise}.  
 
 The obvious disadvantage of the pulsed approach is that high intensity fields only persist for a short period of time in any particular location, and over a very small volume. This requires a detection technology with a high temporal and spatial resolution so as not to average the signal away.   We propose a novel synchronous detection \tec\ that satisfies both of these requirements. 




Our detection system is a modification of a previously reported technique which is capable of measuring  birefringence with extremely high precision~\cite{hall}.  The basis of the original technique 
  is to frequency lock two continuous-wave (cw), orthogonally-polarized lasers to the same longitudinal mode of a resonator using the Pound-Drever-Hall technique~\cite{pdh,black,hils}.  Our proposal is to use a laser pulse stream from a mode-locked laser  rather than cw lasers to excite the resonator.  The advantages of this approach will become apparent below, but we commence with a description of the cw device as this is valid for both cases.

The fractional frequency difference between the two orthogonally-polarized lasers  is equal to the fractional difference in the optical length of the resonator as measured in the two polarization states:
\eq{fracfreqdiff}{\frac{\fperp-\fpar}{\nu_0} = \frac{\lpar-\lperp}{l_0}.}
where $\nu_0$ is the average frequency of the two modes and $l_0$ is the average length of the resonator.
 A path length difference will arise from any birefringence in the cavity in addition to that coming from any intrinsic birefringence of the cavity mirror coatings~\cite{hall}:
 \eq{fracbirefringence}{\fperp-\fpar  \sim   \frac{ \npar-\nperp}{n_{0}} \nu_0 +\frac{c}{2 n_0 L}  \frac{\delta \phi}{2 \pi}.}
where ${c}/({2 n_0 L})$ is the longitudinal mode spacing of the resonant cavity, $\delta \phi$ is the difference in the reflection phase for the two polarisations, and $n_0$ is the average refractive index in  the resonator.  The laser frequency difference, $\fperp-\fpar$, can be extracted by detecting the beat-note between the lasers  and measuring its frequency with   a conventional high precision frequency counter. 


We note that cavity length fluctuations that arise from vibration or temperature fluctuations will be common to both polarizations and hence do not appear in the measured frequency difference signal.  This avoids the need for high quality vibration isolation or temperature control of the detection  resonator. 
 
A number of previous experiments have shown that   over a certain frequency band it is feasible to suppress all technical noise sources that afflict frequency locking systems~\cite{day,bondu,uehara}.  In this case the residual frequency instability is limited by the inherent quantum noise of the detected light itself (shot noise).  
%
%
A simple estimate shows that this will limit the accuracy of each locked laser frequency to~\cite{hall,black}:  \eq{dshot}{\delta_{\mathrm{shot}} \sim \frac{c}{2LF} \sqrt{\frac{h\nu}{\Pd \ti}}.}
where $h$ is Planck's constant, $\nu$ is the laser frequency, \Pd\ is the power falling on the feedback photodiode,  $\ti$ is the integration time, and  where we use  an optical resonator of length $L$ with finesse $F$. An experiment using 800\,nm laser light and a detected optical power of a few milliwatts allows stabilization of the  laser  to $10^{-8}$ of the  cavity bandwidth after 1\,s of integration time.   For a measurement of the difference between two mode frequencies, the expected sensitivity is equal to the residual frequency instability of one laser multiplied by $\sqrt{2}$ (because a comparison is being made between two uncorrelated and equally noisy signals).  
Thus with 800\,nm lasers we get a fractional frequency (or length) sensitivity of:
\eq{Ndnurel}{\delta\nu_\mathrm{rel}&\approx&6\times 10^{-19}\sqrt{\frac{1\unit{s}}{\ti}}\brac{\frac{4\unit{m}}{L}}\brac{\frac{10^5}{F}}\sqrt{\frac{5\unit{mW}}{\Pd}}}
In order to access this level of sensitivity we modulate the expected \vb\ effect so that  the useful signal falls in the shot-noise limited part of the sensitivity spectrum.


To generate \vb\  we use an auxiliary linearly polarized `pump' laser beam to interact with the two detection beams.  In order to exploit the polarization dependence of the coupling between the pump and orthogonally-polarized detection beams we align the  pump field polarisation with the polarization of one of the detection beams.  In order to maximize the interaction between the detection and pump beams they need to be exactly counter-propagating and coaxial~\cite{alek}. 
Unfortunately, coaxial pump and detection beams will overlap on the resonator mirrors and it has been shown that a photo-refractive interaction between the beams in the mirror coatings can generate spurious birefringence signals~\cite{hall}. 
%
Thus we use a second optical resonator to enhance the power of the pump beam, as illustrated in Fig.~\ref{dualcavities}, which lies at an angle, $\theta$,  to the axis of the detection resonator.    The resonators are of identical length $L$, and the separation between the resonator axes of $x$ at the cavity mirrors of radius, $a$. 
\makefig{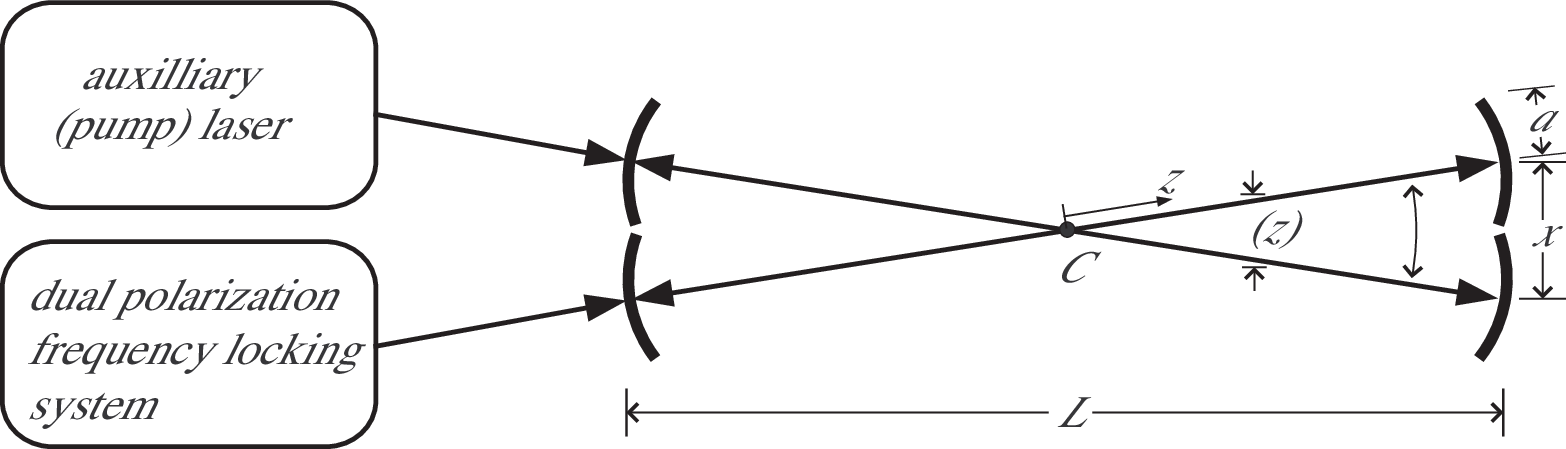}{dualcavities}{Measurement scheme for optically-induced birefringence.}  
The crossed cavity design potentially leads to a limited interaction zone between the pump and detection beams. 
A  key suggestion of this letter is that if both detection and pump beams are pulsed  rather than continuous-wave (cw) signals and are synchronized so that the detection and pump pulses meet head on at $C$ (see Fig.~\ref{dualcavities})~\cite{ma,joneshol}, and in addition, each of the pulses is short enough to completely pass through the other before the beam axes begin to separate, then all of the light circulating in the detection cavity will interact with all of the light circulating in the pump cavity on every pass.  Furthermore, the pulses interact   while the beams are tightly focussed, and therefore where they are most intense.  
  Mode-locked lasers have already been frequency-locked to a resonator with relatively high precision~\cite{jones-lock}, and low-dispersion  cavities can allow even very short pulses to be coupled into high-finesse resonators with low power loss and relatively little broadening of the circulating pulse with respect to the input pulse~\cite{jones-dco,me-pulse}. 


To determine the sensitivity of this approach we 
calculate the the average intensity seen by a pulse circulating in the detection cavity, which is equal to
		\eq{Iav}{\Iav&=&\frac{1}{L}\Lint{I(z)}}
where $I(z)$ is the intensity as a function of longitudinal position in the cavity.  When short pulses are used, it is only  the small region approximately half the pulse length either side of the beam crossing point that contributes significantly to the above integral.  
If  the separation, $\rho(z)$, between the beam axes remains significantly less than the beam radii in the interaction region, then the beams can be treated as approximately coaxial when calculating \Iav.  

The minimum angle by which the beams must be skewed is set by the mirror spacing, $x$, 
 which is determined by the size of  the cavity mirrors. The mirror radius, $a$,  is set by the need for the mirrors to be larger than the laser spot radius evaluated at the mirror position, $w(L/2)$, by a factor $\alpha$. The value of $\alpha$  is determined by the extent to which aperture losses can be tolerated for a particular application. 
In the limit of a small beam waist size, $w_0$,  the spot size at the mirrors can be calculated as
 		${x\approx  {\alpha L\lam}/{\pi w_0}}$.
	Using these results we have shown elsewhere~\cite{andjess} that 
	 the beam waist has to obey the following inequality:
	\eq{wo1}{
 			w_0\gtrsim 10\micron
 			\sqrt{\frac{\tau}{200\mathrm{\,fs}}}^{\frac{1}{2}}
 			\brac{\frac{\alpha}{4}}^{\frac{1}{2}}
 			\brac{\frac{\lam}{800\mathrm{\,nm}}}^{\frac{1}{2}}.}
	to ensure  that the beams do not significantly separate within the interaction zone.  In addition, under this same limit  the Rayleigh range of the interacting beams is significantly  longer than the spatial extent of the pulses and thus the beams are also essentially collimated in the interaction zone. Under these conditions the detection beam sees all of the pump energy on each pass through the interaction zone.

The circulating pump pulse energy, \Ep\, is determined by the average input power \Pav , the repetition rate, $R$, of the input pulse train, resonator finesse $F$, and an efficiency factor, $k_\mathrm{cav}$, which allows for imperfect mode matching, impedance matching and dispersion related losses~\cite{me-pulse}:
\eq{ep}{\Ep=k_\mathrm{cav} \frac{F}{\pi}\frac{\Pav}{R}.}
Since we wish the circulating pulse to be efficiently reinforced on each pass by the incident pulse train, the free spectral range of the cavity must be identical to the repetition rate of the laser, $R$~\cite{jones-dco,me-pulse}.
In this case we find, that:
\eq{Iav3}{\Iav\apx\frac{2 k_\mathrm{cav}\log 2 }{(\pi w_0)^2}F\Pav.}
	
	Combining the expressions above, we arrive at   the following indicative numerical expression for \Iav.
\eq{Iav4}{\Iav\apx \frac{F}{52\, \mathrm{x} 10^{3}}\frac{\Pav}{20\unit{W}}\frac{200\unit{fs}}{\tau}\frac{4}{\alpha}\frac{k_\mathrm{cav}}{1}\frac{800\unit{nm}}{\lam}\times1.5\unit{\frac{PW}{m^2}}}
where we have made use of  achievable experimental parameters.  A finesse of 52\,000 corresponds to a reflectance of $99.994\%$ which is available in a custom low dispersion mirror coating~\cite{ltech}, that has sufficiently low dispersion to allow 200\,fs incident laser pulses to be directly coupled into a cavity with near-unity efficiency~\cite{jones-dco,me-pulse}.  A mode-locked laser has been reported with a 200\,fs duration output pulse and  20\,W  average power~\cite{brunner}.
Thus, using readily available equipment  it should be possible to construct a pump cavity which gives a effective average  intensity in the detection cavity of  1.5\,$\mathrm{PW/m}^2$.  Such high average intensity is only possible because  the arrangement of pulsed and counter-propagating detection and pump beams circumvents the high divergence that would normally afflict tightly focussed beams. In fact, the pulsed beams show the same degree of  interaction as  cw beams that were parallel and nondivergent throughout the cavity, which is of course impossible for tightly focussed, non-coaxial beams.

	 The optimal cavity length in a real experiment relies on constructing sufficiently large mirrors as implied by the beam size calculation above.  For a 3\,m  cavity with $\alpha=4$, the mirrors would need to be 20\,cm in diameter.  Although this presents a significant challenge, it is not insurmountable, as demonstrated by the recent construction of even larger diameter mirrors for gravitational wave interferometers~\cite{mirrors}.

	A very important  advantage of our particular approach  is the ability to modulate the effective strength of the interacting fields without varying the energy load or distribution on the mirror surfaces.  This enables detection of the birefringence signal in a frequency domain where there is minimal noise interference without giving rise to spurious signals. We achieve this effective power modulation by temporally delaying or advancing the pump pulse with respect to the detection pulse and thus varying the degree of energy overlap at the crossing point of the two cavities.  The absence of modulation  in the thermal load on the mirrors  eliminates many potential spurious effects that could otherwise masquerade as the effect of interest.  This technique can be implemented as part of the control system that synchronizes the detection and pump pulses~\cite{ma,joneshol}.

		

Various challenging technical issues must be addressed in order to implement this experiment. For example, the pump and detection beams must be synchronized so that the pulses meet where their axes cross~\cite{ma,joneshol}. This implies that the offset frequency and repetition rate of the pulse trains from the  lasers must be controlled so as to match the cavity resonance frequencies and free spectral range of both resonators~\cite{jones-lock}.  The final hurdle to the detection of vacuum birefringence will be the duration of the experiment observation time in order to unambiguously detect  the effect.  We calculate these integration times by equating the expression for shot-noise limited measurement sensitivity in Eq.~(\ref{Ndnurel}) and the expected vacuum birefringence signal in Eq.~(\ref{b})  and present them in Table~\ref{tab1}. 
 The first two lines predict the performance available using  low dispersion mirrors: the first line shows the capability of the best ``off-the-shelf'' commercially available low dispersion mirrors while the second line shows the capability of the best custom built mirrors~\cite{ltech}.  The final line of the table predicts that performance that could be achieved  if   low dispersion mirrors were to have a reflectivity equal to that of the best commercially-available super-mirrors. 


\begin{table}[tb]
\centering
\begin{tabular}{|c|c|c|}
\hline
$R$, \%&$F$&$\tau_\mathrm{int}$\\
\hline
99.97&1.0\ee{4}&2.6 years\\
99.994&5.2\ee{4}&1.7 days\\
99.997&1.0\ee{5}&2.5 hours\\
\hline

\end{tabular}
\caption{Required integration times  for the detection of vacuum birefringence as a function of resonator mirror reflectivity and resonator finesse, $F$.  The other experimental parameters are explained in the text.}
\label{tab1}
\end{table}
  
 The measurement time required to detect \vb\ scales with the inverse fourth power of the finesse since the finesse affects both the measurement sensitivity and the average intensity in our proposal approach.  Competing techniques that rely on a macroscopic magnetic field to create the vacuum polarization have  an integration period that decreases as the square of the finesse of the detection cavity. 
  If low dispersion mirrors could be improved to the point that 99.997\% reflectivity mirrors became available (only as good as existing super-mirrors) then  the corresponding increase in finesse would allow vacuum birefringence to be detected in just a few hours.  Any increases in  available laser power would also reduce the required measurement time.
 


We have proposed a completely new approach to the experimental detection of very low levels of field-induced birefringence.   The system is based on the intersection of two  high-finesse short-pulse resonant cavities, one of which pumps the QED vacuum to produce the birefringence, while the other detects this induced birefringence using highly sensitive frequency metrology techniques. Current limits in mode-locked laser technology and low-dispersion mirrors  already allow detection of the predicted vacuum nonlinearity with a measurement period of just a few days. Readily forseeable advances in laser output power and mirror coating technology should reduce the required integration time for such an experiment to just hours. This approach has eliminated many of the defects that limit the performance of alternative techniques that use superconducting magnet systems.
 
 \section{Acknowledgements}
 
 We thank the Australian Research Council (ARC) for financial support of this research. We acknowledge all members of the Frequency Standards and Metrology Group  which make it such a pleasant and stimulating environment in which to work. 


\end{document}